\documentclass[a4paper,11pt]{article}
\pdfoutput=1 

\usepackage{jcappub} 

\usepackage[T1]{fontenc} 

\usepackage{graphicx}
\usepackage{dcolumn}
\usepackage{bm}

\usepackage{hyperref, color}
\usepackage{amsmath,braket}
\hypersetup{colorlinks=true, linkcolor=magenta, filecolor=cyan, urlcolor=blue}

\def\@Aboxed#1&#2\ENDDNE{%
  \settowidth\@tempdima{$\displaystyle#1{}$}%
  \addtolength\@tempdima{\fboxsep}%
  \addtolength\@tempdima{\fboxrule}%
  \global\@tempdima=\@tempdima
  \kern\@tempdima
  &
  \kern-\@tempdima
  \fcolorbox{red}{yellow}{$\displaystyle #1#2$}
}

\newlength\dlf

\newcommand{\be}{\begin{equation}}
\newcommand{\ee}{\end{equation}}
\newcommand{\bea}{\begin{eqnarray}}
\newcommand{\eea}{\end{eqnarray}}
\newcommand{\bma}{\left(\begin{array}}
\newcommand{\ema}{\end{array}\right)}

\usepackage{float}

\title{\boldmath Inflation and Electroweak Symmetry Breaking}

\author[1]{Stephon Alexander}

\affiliation[1]{Brown Theoretical Physics Center and Department of Physics, Brown University,\protect\\  Providence, RI 02912, USA}

\emailAdd{stephon\_alexander@brown.edu}

\abstract{We present a model of Cosmological Electroweak Symmetry Breaking (CEWSB), where a Higgs-like field and a cosmological background of weak boson gauge fields interact with gravity to realize the epoch of cosmic inflation, which is then followed by a Higgs resonance preheating.  As a result, the scale of electroweak symmetry breaking is linked with the end of inflation.  The theory is equipped with a shift symmetry that can protect the Higgs mass, and it has close semblance to natural inflation and its variants.  As the Higgs field's amplitude decays at the end of inflation, its mass emerges. The model has a built in Higgs self-resonance preheating mechanism which leads to the possible emergence of the cosmic microwave background (CMB) due to resonant Higgs, quark and lepton production after inflation.  We provide a pathway to implement a similar mechanism with the realistic Higgs-doublet of the standard electroweak theory and discuss phenomenological considerations.}

\begin{document}
\maketitle
\flushbottom

\section{\label{sec:background}Introduction}

The discovery of the Higgs boson, with a mass of 125 GeV, has sharpened questions about its quadratic sensitivity to high scales due to quantum corrections \cite{lhcresult}. Supersymmetry and other approaches provided new dynamics, with the use of extra states, to tame these divergences.  So far, no new exotic states have been found that could address the the Higgs hierarchy problem. Since the Higgs field is the only observed candidate of a fundamental scalar field, it is natural to ask if it can actually be the inflaton field, and if the early universe can address the Higgs Hierarchy problem \cite{guth}. 

However, if cosmic inflation were to be driven by the standard Higgs potential, cosmic microwave background (CMB) constraints would force the Higgs self coupling to be $\lambda \sim 10^{-13}$, which is inconsistent with the LHC measurent of $\lambda \sim 0.13$ \cite{Linde:1983gd}. Improvements in Higgs-inflation were made in pioneering work by Bezrukov and Shapshnikov by introducing a quadratic Higgs coupling to the Ricci scalar \cite{BS}.  These models make robust predictions for the CMB and posit some issues regarding the breakdown of unitarity in the Higgs EFT \cite{burgess,Hertzberg:2010dc}.  In what follows, we provide a new mechanism that does not require non-minimal curvature couplings.  In this work we ask the question: Does the scale of EWSB result from the termination of inflation?

With inspiration from shift symmetry, we will first consider a toy model which comprises of a Higgs field as an $SU(2)$ triplet, as we will see that the quantitative analysis is in direct correspondence with natural inflationary models based on Chern-Simons couplings \cite{freese,ams,Adshead,Maleknejad}. The minimal Higgs triplet considered in this paper is no longer a good candidate for the observed Higgs doublet, and it is not meant as a replacement for the Standard Model Higgs doublet\footnote{These triplet extensions of the Standard Model are interesting in their own right as it can resolve a handful of problems such as neutrino masses and others \cite{Schechter:1981cv}.}. It is possible to also consider a dark $SU(2)$ gauge and Higgs sector, and this triplet model could function in that sector. Such models are useful in providing new ideas of dark matter in the form of light mesons \cite{aes,aes2,mal}.  However, many ingredients of this model will carry over to a more realistic Higgs doublet, and we will discuss this avenue for discussion at the conclusion and for future work.
\newline
\indent
For clarity, it is instructive to summarize this mechanism before delving into the details.  Inflation begins in a spatially homogeneous flux configuration of weak bosons and a Higgs condensate that spontaneously break $SU(2)_{W}$.  The homogeneous Higgs field is necessarily time-dependent and rolls down a local minima of its periodic potential.  At the end of inflation, the Higgs mass emerges as a result of the decay of its time-dependent amplitude.  As the Higgs oscillates around its minimum, it generates a parametric amplification of Higgs particles and weak bosons. 
\newline
\indent
In section II, we present the new Higgs sector and discuss its new features and collider constraints on new interactions. In section III, we discuss the embedding of this model into Chern-Simons/chromo-natural inflationary models.  In section IV, we explore a self-resonance Higgs preheating mechanism inherent to the model.

\section{The Theory}

To construct a model that both protects the Higgs mass and generates inflation driven by the Higgs field, we introduce a periodic potential for the Higgs field which has a shift symmetry, $H_{a}\tau_{a} \rightarrow H_{a}\tau_{a} + 2\pi f$, prohibiting an explicit mass term\footnote{Periodic potentials, in an axionic sector, have been implemented in the Relaxion mechanism to address the Higgs naturalness in the context of inflation\cite{relax}.}. Here $H_{a}$ is the Higgs field and $\tau^{a}$ are the generators of the $SU(2)$ gauge transformation as $H\rightarrow U(x)H U^{\dagger}(x), U(x)\in SU(2)$.  

The Higgs field has a periodic potential similar to what is obtained in models where the Higgs is a PNGB \cite{Georgi,Arkani-Hamed,Croon:2015fza}. In this picture of compositness, the Higgs is similar to pions in QCD, where some other strongly interacting theory triggers the formation of bound states of fermions \cite{steve,lenny,Georgi,Arkani-Hamed}. It is possible that a pre-inflationary phase is linked to emergence of the Higgs.  \footnote{It is interesting that a Yang-Mills theory based on the spin connection of General Relativity can exhibit a negative beta function, which may provide a mechanism of a gravitational formation of a composite Higgs due to the pairing of fermions, and we will pursue this in the future \cite{Donoghue,nesti,iso}.}

With these ingredients, we seek to generate inflation, reheating and EWSB from a minimal EFT that only involves general relativity, the Higgs and weak vector bosons. Hence, the action becomes\footnote{From an EFT perspective, since we have a dimension five Higgs-W-boson operator in our action, we can wonder if a dimension five $HFF$ operator can appear. However since we are dealing with a Higgs Triplet, there is no gauge invariant way to make this operator. It is possible to have a dimension six operator $H^{2}FF$ and these LHC constraints have been considered by the authors \cite{Pomarol:2013zra}. Note that these operators are built from a Higgs doublet, while we are dealing with a triplet, so we intend to discuss such constraints in an upcoming work that deals with both the Higgs doublet and triplet.}: 

\begin{align}
    S &= \int d^{4}x \sqrt{-g} \bigg[M_{Pl}^{2}R + (D_\mu H)^{\dagger}(D^\mu H) -\frac{1}{2} \text{Tr}[{W}_{\mu\nu} {W}^{\mu\nu}] \nonumber \\ &+ \mu^{4} \left[\cos\left(\frac{H_{a}\tau_{a}}{f}\right)+1\right] + \beta\epsilon^{abc}H_{a}W_{b\mu\nu}\tilde{W}^{\mu\nu }_{c} \bigg], \label{eq:action}
\end{align}
where $W_{\mu\nu}$ is the weak-field strength tensor and the covariant derivative acting on the Higgs is $D_\mu = \partial_{\mu} - ig\tau^{a}W^{a}$.  As we will see, the Higgs-Chern-Simons coupling, $L_{CS} =\beta\epsilon^{abc}H_{a}W_{b\mu\nu}\tilde{W}^{\mu\nu }_{c}$, is motivated from Chromo-Natural inflation, so as to add extra friction to the steep cosine potential to achieve a viable inflationary mechanism.  Here the coupling to the Chern-Simons term $\beta$ has dimension $[M]^{-1}$. The shift symmetry acting on the Chern-Simons term induces a total derivative ${\rm Tr} [W\tilde{W}]$, leaving the action invariant under the discrete shift symmetry.

A caveat is that the Higgs kinetic term is not shift-symmetric.  However, in the high-energy regime, the gauge theory is massless, and if the theory is asymptoically free, then the shift symmetry will be softly broken. It is possible to embed this in a minimal Little Higgs model where the weak group is extended to $SU(3)$; in this case the theory will be asympotically free \cite{sch1,sch2}.  We will pursue this in an upcoming work\footnote{At this point, this is a model-agnostic statement; such shift symmetries could arise from any model that treats the Higgs as a pseudo-Nambu-Goldstone boson, e.g., Little Higgs or other such models.  Also, the Higgs kinetic terms break the shift symmetry, but if the gauge couplings are small, the shift symmetry will be softly broken.  We thank Devin Walker for pointing this out.}. 

The pattern of $SU(2)$ symmetry breaking is different from the usual spontaneous symmetry breaking in the standard electroweak theory.  As we will demonstrate later, unlike that of global Minkowski spacetime, our Higgs field amplitude evolves in FRW spacetime and is necessarily time dependent-suggesting the EWSB is also time dependent.  Initially, the Higgs field sits at the top of the periodic potential and will be homogeneous and time dependent, $H(t,x) = H(t) + h(x,t)$, initiating a period of slow-roll inflation. As the Higgs field slow-rolls during inflation, it spontaneously picks a direction in $SU(2)$ color space.  In what follows, we will first explore conditions to obtain a viable epoch of inflation and reheating with this model.  

\section{Higgs-W-Boson and Chromo-Natural Inflation}
The extended Higgs Lagrangian, (\ref{eq:action}), has close semblance to that found in Chromo-natural and Chern-Simons inflationary models. In those models, inflation is driven by a pseudo-scalar field coupled to a dark $SU(2)$ gauge field \cite{ams,Martinec:2012bv,Adshead,Maleknejad}. In our case, the inflationary scalar is the Higgs field of the Standard Model.  The role of the gauge field will be to flatten the steep Higgs potential.  As a result, the field equations will differ mainly due to the extra $SU(2)$ coupling of the Higgs field to the gauge bosons, compared to a singlet scalar. This difference is crucial as it determines the nature of EWSB in cosmology. In what follows, since the Hubble parameter is typically denoted $H$, we will rename the Higgs field $H \rightarrow \Phi_{H}$.

It is important to note that the original version of Chromo-Natural Inflation, while providing the correct background inflationary dynamics, is ruled out due to overproduction of gravitational waves.  However, it has been shown that if the gauge symmetry is sponataneously broken, Chromo-Natural Inflation can be made consistent with CMB observations \cite{wa}. Our model is a spontaneously broken gauge theory, and it would be necessary to revisit this model in a future work to see if the CMB observables such as $n_s$ and $r$ and consistent with recent CMB data.

Inflation is a regime whose initial condition corresponds to a field that is isotropic and homogeneous.  We postulate that at the beginning of inflation, not only does the Higgs develop a spatialy homogeneous condensate $H(t)^{a}$, but it also spontaneously picks a direction in color space.   This is consistent with the fact that $SU(2)$ invariant gauge fields are compatible with isometries of the FRW metric if the color indices are locked with the spatial vector indices.  The homogeneity and isotropy spontaneously breaks the electroweak symmetry. Since the Higgs field is a triplet, in what follows we assume that it couples to a background dark $SU(2)_{R}$ gauge field. This connects nicely to past work that couples axions to $SU(2)_{R}$ that has rich phenomology to dark matter and baryogenesis \cite{Maleknejad:2020yys}. It is possible that if the Higgs triplet couples to a background isotropic $SU(2)_{L}$ W-boson field, constraints on isocurvature perturbations may render the standard model coupling inconsistent\cite{Papageorgiou:2018rfx}.  This requires a careful perturbation analysis that also includes the Higgs doublet and we leave this for future work and assume that the background gauge field is a dark $SU(2)_{R}$ sector \footnote{We thank the referee for pointing out the possible inconsistencies in CMB physics if the standard model Weak Bosons are background time dependent cosmological fields}.   Once the Higgs picks a VEV at the beginning of inflation, it will roll down the bottom of the cosine potential.

Inflationary slow-roll trajectories in the above action (\ref{eq:action}) are found by considering the Higgs  $\Braket{\Phi_{H}}^{a} =\Phi(t)_{H}$ in a homogeneous configuration and the gauge flux in a classical time-dependent configuration:

\begin{align}
W_{0}^a = & 0, \quad W_{i}^a =  aQ(t)\delta^{a}_{i}, \label{eq:a-vev}
\end{align}
where $\tau_a$ is the generator of $SU(2)$ and $a$ is the scale factor.  Likewise the field strength tensor becomes:
\begin{align}
W_{0i} = & \partial_{\tau}Q \delta^{a}{}_{i}t^a,\quad
W_{ij}  =  g Q^2 f^{a}_{ij} t^a.
\end{align}

One may worry that identifying the Higgs and weak bosons as the drivers of cosmic inflation would cause phenomenological issues. For example, after inflation the Universe could be populated with potentially large amounts of background Higgs and vector bosons.  But as we will soon see, the rapidly expanding background of inflation will eventually dilute the weak bosons that were active during inflation.  As for the Higgs condensate, we will see that at the end of inflation, it will decay into massive Higgs particles through parametric resonance during the preheating epoch. 

In order to see if inflation is possible in this system, we must see if the combined FRW equations and Higgs-Gauge equations of motion satisfy the slow roll conditions, and if the parameters are consistent with at least 60 $e$-foldings. Thus the Friedmann equations governing the evolution of system are
\begin{align}
\label{H2}
3H^2 M_p^2  = \frac{1}{2}\dot \Phi_{H}^2 + V(\Phi_{H}) + \frac{3}{2} \left[  \left( \dot Q + HQ \right)^2+ g^2 Q^4 \right]
\end{align}
and the second Friedmann equation is

\be
 \dot{H}=-\left(\frac{\dot{Q}^2}{a^2}+\frac{g^2 Q^4}{a^4}+\frac12\dot\Phi_{H}^2\right).
\ee
\DeclareGraphicsRule{}{}{}{}

The slow roll conditions are
\be \label{slow-roll-par}
\epsilon \sim \frac{\dot H}{H^2},\quad \eta \sim \frac{\ddot{H}}{2\dot{H}H}.
\ee

Inflation occurs when we have the slow roll conditions $\epsilon \ll 1$ and $\eta \ll 1$, which imply that the  potential $V \left( \Phi_{H} \right)$ is the dominant contribution to inflation:

\be 3H^2 M_p^2 \sim  \mu^{4}\cos\left(\frac{\Phi_{H}}{f}\right) \ee

The Higgs-Gauge equations of motion couple to the above Einstein equations and their dynamics have to sustain the slow-roll conditions in order to have enough $e$-foldings.  Their equations of motion are:

\be \label{axionEq}
\ddot{\Phi}_H + 3 H\dot{\Phi}_H + g^{2}W^{2}\Phi_{H} + V'(\Phi_H) \, = \, \beta{\vec{E}_{a}} \cdot {\vec{B}_{a}} \, ,
\ee

\bea
\partial_{\mu}W^{\mu\nu} - 4\beta \Phi_{H a}\epsilon^{abc}A_{b\mu}\Tilde{W}_c^{\mu\nu} = 0.
\eea

\bea
\ddot{\Phi}_{H}+3H\dot{\Phi}_{H} +g^{2}\Phi_{H}Q^{2}-\frac{\mu^{4}}{f}\sin\left(\frac{\Phi_{H}}{f}\right)=-3g\beta Q^2\left(\dot{Q}+HQ\right),\label{Higgseom}
\eea

\be
\ddot{Q} + 2H\dot{Q} + (\dot{H} + H^2)Q = 8g\beta \Phi_H Q^2a^2.
\ee

In recent work, a mass term for the gauge field was added to Chromo-Natural Inflation, which rendered the isocurvature perturbations to be negligible, allowing resulting adiabatic fluctuations to match current observational constraints\cite{ADShiggs}. These authors also demonstrated that the slow roll trajectories will persist with the inclusion of the mass term.  In this model, the Higgs induces a mass term naturally, as can be seen in the third term in eq\eqref{Higgseom}.  Hence, the slow-roll conditions implemented in \cite{ADShiggs} will carry over to this case.  When we apply the slow-roll conditions, the W-boson field ends up in an attractor configuration \cite{Adshead}:

\begin{eqnarray}
Q_{\rm min}= \left(\frac{\mu^{4}\sin(\Phi_{H}/f)}{3g\beta f H}\right)^{1/3}.
\label{Qmin}
\end{eqnarray}
This applies for inflation that takes place when the initial Higgs field $\frac{\Phi_{H_{0}}}{f}$  starts at the top of the cosine potential and rolls down to the local minimum, $\frac{\Phi_{H}}{f} \sim \pi + O(\epsilon)$.  
Finally, we can compute the number of e-foldings, using it as a time variable $N=Hdt$ given by the following integral:
\be\label{N}
N\simeq\frac{\beta f\Omega}{\sqrt[3]{4}}\int^\pi _{\frac{\Phi_0}{f}}\frac{(1+\cos x)^{2/3}(\sin x)^{1/3}}{\Omega^2(1+\cos x)^{4/3}+(2\sin x)^{2/3}}dx,
\ee
where $x\equiv\frac{\Phi_{H}}{f}$ and $\Omega\equiv(\frac{2\beta f\tilde{\mu}^4}{3g^2})^\frac{1}{3}$.  Where the dimensionless Higgs potential height is $\tilde{\mu}=\frac{\mu}{M_{Pl}}$.
Demanding 60 $e$-foldings yields the dimensionless coupling $\beta f \sim 100$, which is consistent with the effective field theory \cite{mal}. Moreover, perturbations in chromo-natural inflationary models, as well as observable windows for detectable gravitational waves, have been explored \cite{Dimastrogiovanni:2012ew,Adshead:2013qp}.

\vskip 0.5 cm

\section{Self Resonance of the Higgs and Preheating\label{sec:3}}
One of the most important aspects of any inflationary model is how to create the CMB, as well as the plasma of quarks and leptons by the epoch of last scattering.  In any model of inflation, the Inflaton field must dump its energy into photons, leptons and quarks to form the CMB by last scattering.  This falls into the category of reheating or preheating and introduces another level of fine-tuning: the inflaton's coupling to the Standard Model.  

In our model of Higgs inflation, we may not have this fine-tuning problem, since the inflaton itself is the Higgs field.  Because of gauge invariance, the Higgs field already couples unambiguously to the matter fields of the Standard Model. While there are a handful of possible channels of reheating, both perturbative and non-perturbative, we will present the possibility that the homogeneous part of the Higgs field, which drives inflation, can play a dual role of self-resonating the Higgs particle at the end of inflation, leading to an explosive production of Higgs bosons through the phenomenon of parametric resonance.

As the background Higgs condensate slow rolls down its inflationary potential, its field amplitude decreases, the condensate approaches the local minimum of the periodic potential, and inflation comes to an end. At this stage, the field oscillates about the minimum with a frequency proportional to the curvature of the potential:
\begin{align}
    \Phi_{H}(x,t) = \Phi(t)_{H} + h(x,t).
\end{align}

The background evolution of the time-dependent homogeneous Higgs field is 
\begin{align}
    \ddot{\Phi}(t)_{H} + 3H\dot{\Phi}(t)_{H} + M_{H}^{2}\Phi(t)_{H}=0, 
\end{align}
whose general solution is a decaying oscillatory behavior, $\Phi_{H} = A\text{e}^{-\alpha t}\sin(\omega_{M}t)$.

We likewise obtain the equation of motion for the Higgs field perturbation $h(x,t)$:
\begin{align}
    \ddot{h}(x,t) + 3H\dot{h}(x,t) + \nabla^{2}{h(x,t)} = A\sin(\omega_{M}t).
\end{align}
We can do a Fourier mode decomposition

\begin{equation}
h \left( t ,\, {\bf x} \right) = \int \frac{d^3 k}{\left( 2 \pi \right)^{3/2}} {\rm e}^{i {\bf k} \cdot {\bf x}} \, h_{k},
\end{equation}
and neglecting expansion, the Higgs mode equation becomes\footnote{We ignore expansion because we are in the regime where the slow-roll condition is violated and inflation ends, and the dominant contribution to the energy density is the kinetic energy of the Higgs field.}
\begin{align}
    \ddot{h}_{k}(t) + \omega(k,t)^{2}h_{k}(t) =0,
\end{align}
where $\omega^{2}(k,t) = k^{2} + A\cos(mt)$ and the period of oscillation is $T=\frac{2\pi}{m}$.

This equation is the well-known Mathieu equation and leads to a parametric amplication of the Higgs field characterized by both narrow and broad bands of wavevectors.  This system has been studied in considerable detail and we refer the reader to \cite{Shtanov:1994ce,Kofman}. We will only discuss the general features of Higgs preheating relevant to our model.  The Floquet theorem states that the most general solution to the Mathieu equation is 
\begin{align}
    h_{k}(t) = \text{e}^{\mu_{k}t}P_{k+}(t) + \text{e}^{-\mu_{k}t}P_{k-}(t),
\end{align}
where $\mu_{k}$ is the Floquet exponent and $P_{k\pm}(t) =P_{k\pm}(t+T)$.

The energy per mode is 
\begin{align}
    E_{k} =\frac{1}{2\pi^{3}}\bigg(n_{k}^{h} + \frac{1}{2}\bigg)\omega(k,t),
\end{align}
where $n_{k}^{h}$ is the Higgs particle mean occupation number.  For the first band of  parametric amplication, the occupation number is exponential, $n_{k}^{h} \sim \exp(At/m)$.

On a similar footing, the $W$-boson perturbations will also be active.  The vector perturbations couple to the metric perturbations. In the scalar  longitudinal gauge  and  vector  gauge,  the most  general
perturbed FRW metric is

\begin{equation}
\begin{split}
   ds^2 =a^2(\eta)[-(1+&2\Phi)d\eta^2 - 2 B_i\, d\eta dx^i
    \, \\ & + ((1 + 2\psi) \delta_{ij} + 2 h_{ij} )dx^i dx^j ],
\end{split}
\label{eq:perturbedmetric}
\end{equation}
where $B^i$ is the transverse metric vector perturbation, $\partial_i B^i=0$, and $h_{ij}$
is a transverse and traceless tensor perturbation, $\partial_i h^i{}_j=h^i{}_i=0$.
The  equations  of  motion   for  vector  (metric)  perturbations  are
\cite{Wayne,Kodama,ams}
\begin{equation}\label{eq:vectormetric}
  \partial_j B^i{}'+2 \frac{a'}{a}\partial_j B^i{}
  =8\pi G a^2 p_A {}^{(v)}\Pi^i{}_j,
\end{equation}
where $^{(v)}\Pi^i{}_j$  is the vector  part of the  W-boson anisotropic
stress tensor.  Inflation will redshift away the anisotropic stress sources, and Eq.
(\ref{eq:vectormetric}) dictates that the vector perturbations decay away,
$B^i \propto 1/a^2$. So even if they are generated during inflation,
they will quickly redshift at later times.  We can now focus on the perturbations of the $W$-bosons; the density of $W$-bosons will redshift as $\frac{1}{a^{4}}$. 

We now want to consider the dynamics of the weak boson perturbations after inflation ends, so we expand the gauge fields as:

\begin{align}
 W^a_i(t, \vec x) & = \tilde{Q}(t) \, \delta^a_i + \delta W^a_i(t, \vec x) \,, 
\end{align}
where $\delta W^a_i(t, \vec x)$ are the gauge fluctuations around the homogeneous background. After some considerable algebra, expanding the Lagrangian up to quadratic order in the gauge field fluctuations and linear order in the gauge coupling, we get the gauge field perturbation equations of motion:

\begin{align}
& \delta W_i^{a \prime \prime } - \partial_j \partial_j \delta W_i^{a}  + \partial_i \bigg( -\delta W_0^{a \prime} - \partial_j \delta W^{a}_j \bigg) \nonumber \\
& - g\varepsilon^{a b c } \bigg[ -2 \delta W_0^{b} \delta^c_i \tilde{Q}^\prime  + 2 \tilde{Q}\delta^b_j \partial_j \delta W_i^{c} \nonumber \\ & + \tilde{Q} \delta^c_i  \bigg(- \delta W_0^{b \prime} + \partial_j \delta W^{b}_j\bigg) - \tilde{Q} \delta^b_j \partial_i \delta W_j^{c} \bigg] +\Phi_{H}\delta W_i^{a} \nonumber \\
 & - \beta \bigg[ \Phi_{H}^\prime \varepsilon_{ i j k} \bigg(2 \partial_j W^a_k + 2 g\tilde{Q} \varepsilon^{abc} \delta W_j^{b} \delta^c_k\bigg) \nonumber \\ & + 2 \tilde{Q}^\prime \varepsilon^{aji}\partial_j h + 2 g \tilde{Q}^2 \delta^a_i h' \bigg] = 0. 
\label{eq:eom_i_linear}
\end{align}

This equation requires a combination of analytic and numerical treatment to be pursued in an upcoming work \cite{AH}.  However, to get a feel for what is possible in preheating, as argued earlier, after inflation the background gauge field configuration is vanishing, i.e. $\tilde{Q} \sim 0$.   Also ignoring the temporal gauge configuration, we arrive at the following equation for the $W$-boson fluctuations:

\begin{align}
& \delta W_i^{a \prime \prime } - \partial_j \partial_j \delta W_i^{a}  + \partial_i\partial_j \delta W^{a}_j  \nonumber \\
& +\Phi_{H}^{2}\delta W_i^{a}-  2\beta \Phi_{H}^\prime \varepsilon_{ i j k}  \partial_j \delta W^a_k  =0.
\label{eq:eom_i_linear2}
\end{align}

When the background Higgs condensate is oscillating around the minimum $\Phi_{H} =A \sin(\omega t)$, the resulting weak boson fourier mode fluctuations obey:

\begin{align}
& \delta W_i^{a \prime \prime } - k^{2} \delta W_i^{a}  + k_i k_j \delta W^{a}_j +A^{2}[1-\cos(2\omega t)]\delta W_i^{a} \nonumber \\ & - 2\beta A\omega \cos(\omega t) \varepsilon_{ i j k}  k_j \delta W^a_k  =0.
\label{eq:eom_i_linear2}
\end{align}

We immediately see that this dynamical system is a modification to the Mathieu
equation, with the addition of another periodic driving term from the Chern-Simons coupling that is out of phase 
by $\pi/2  $.  Our main goal in this section is to demonstrate the possibility of parametric resonance, and to leave a more detailed analysis of the possibility to produce a parametric amplification of weak bosons during the preheating phase of the Higgs field to an upcoming work \cite{AH}. 

However, if the Chern-Simons term is subdominant, it is evident that the $W$'s occupation number is exponential for a narrow band in k-space, $n_{k}^{W} \sim \exp(At/m)$.  We expect a similar parametric resonance of the decay of the Higgs into leptons and quarks due to the Yukawa couplings.  As a result, we are to expect fermionic parametric resonance of quarks and leptons, and we leave this up to future work \cite{Greene}.

\section{Towards a realistic Higgs Doublet Model}

So far we have studied a concrete toy model of a Higgs Triplet with a periodic potential and Chern-Simons coupling to the gauge sector.  The theory naturally provides an inflatiionary mechanism as well as a VEV for the Higgs triplet.  Ideally we want to extend this toy model into a realistic Higgs doublet model, which will be explored in an upcoming work. We would like to present some model independent statements about the nature of EWSB in this model and the relevant mass scales.

As stated previously, our Lagrangian does not have any explicit mass term and is replaced by a periodic potential for the Higgs field. The emergence of the mass and VEV is due to a time dependent cosmological evolution of the Higgs inflaton field. During inflation, as the Higgs field rolls down its potential its amplitude decreases and at the end of inflation it will obtain a mass dynamically. We can see this by considering a model independent version of our theory which can also be a realistic Higgs-doublet:  

\begin{align}
    S &= \int d^{4}x \sqrt{-g} \bigg[M_{Pl}^{2}R + (D_\mu H)^{\dagger}(D^\mu H) -\frac{1}{2} \text{Tr}[{W}_{\mu\nu} {W}^{\mu\nu}] \nonumber \\ &+ \mu^{4} \left[\cos\left(\frac{H_{a}\tau_{a}}{f}\right)+1\right]  \bigg], 
\end{align}

The pattern of $SU(2)$ symmetry breaking is different from the usual spontaneous symmetry breaking in the standard Electroweak theory.  Initially an homogeneous and time dependent Higgs field, sits at the top of the Cosine potential at one of the local maxima ($\pi n$) initiating a period of slow roll inflation. As the Higgs field slow-rolls, it spontaneously picks a direction in $SU(2)$ color space.  At late times the Higgs reaches the condition $\frac{H_{a}\tau_{a}}{f} < 1 $ and the cosine potential can be expanded as 
\begin{equation} V(\tilde{H}) = -M_{H}^{2} \tilde{H}^{\dagger}\tilde{H} + \lambda(\tilde{H}^{\dagger} \tilde{H})^{2} \end{equation} 
where $\tilde{H}= H-n\pi f$ and we make the associations
\begin{equation}
    M_H^2 = \left.\frac{dV}{dH}\right]_{H = n\pi f}\qquad\text{and}\qquad\lambda = \left.\frac{d^2V}{dH^2}\right]_{H = n\pi f}
    \label{eq:expansion_terms}
\end{equation}

We can now derive the mass of the Higgs from the two parameters in our theory. Using the expansion in Eq. \ref{eq:expansion_terms} the Higgs mass is identified as $M_{H} =\frac{\mu^{2}}{f}$ and the dimensionless self-coupling is identified as $\lambda=\frac{\mu^{4}}{f^{4}}$. Here we note that the terms in the usual Higgs potential from electroweak theory arise as a result of the shift symmetry of the cosine potential, not as a particular combination of the parameters. This new freedom is one of the salient features of the model. The only two new ingredients of the Higgs sector are the periodic potential and the coupling of the Higgs field to the Chern-Simons Electroweak term. Since this new theory ought to conform to experimental measurement, then $M_H\sim10^2\text{ GeV}$, which constrains the height of the potential, $\mu$, to be of the form $\mu \sim \sqrt{f M_{H}} \simeq 10^{1}f^{1/2}GeV $. We can combine these to find the Higgs self-coupling to be $\lambda \sim \frac{M_{H}^{2}}{f^2}$.  

After EWSB and during reheating, the Higgs field dumps its energy into particle production and obtains its VEV at the conclusion of inflation, $ \Braket{H} = \pi nf$, where $n$ is an integer.  It is possible that different Hubble patches will initiate inflation with different values of $n$ and we leave this possibility up to future work\footnote{We thank Michael Peskin for pointing this possibility out to the author.}. 

Our model nicely ties in the scale of inflation with the Higgs-mass and VEV. Recall that end of inflation coincides with the Higgs mass scale $M_H$, and according slow-roll condition combined with the Friedmann-eq we considered in eq \ref{H2}, this fixes the scale of inflation by $\mu \sim 10^{1}\sqrt{f}$.  The Weak boson mass scales as $m_{W_{\pm}} \sim nf$.  It is interesting that in the $n=1$ vacuum, this scale $\mu$ can coincide with a TeV scale inflationary model which has been advocated for by Knox and Turner some years ago \cite{knox}.

For a more realistic Higgs doublet our theory will change slightly.  While the story of the mass scales discussed above will not be altered, the Higgs doublet model requires us to carefully revisit our cosmological and phenomenological analysis.  Consider the proposed action for the doublet model:

\begin{align}
    S &= \int d^{4}x \sqrt{-g} \bigg[M_{Pl}^{2}R + (D_\mu H)^{\dagger}(D^\mu H) -\frac{1}{2} \text{Tr}[{W}_{\mu\nu} {W}^{\mu\nu}] \nonumber \\ &+ \mu^{4} \left[\cos\left(\frac{H_{a}\tau_{a}}{f}\right)+1\right] + \gamma H^{\dagger}H\text{Tr}[{W}_{\mu\nu} \tilde{{W}}^{\mu\nu}]  \bigg], 
\end{align}

In the above action the only difference will the Higgs-Chern-Simons coupling.  This will impact the inflationary dynamics and provide a new dimension six operator for the Higgs decay into W,Z and photons.  The generic mechanism of EWSB from the triplet action will carry over to the doublet case, but it remains to be seen if the new Chern-Simons term conspires to give a consistent inflationary mechanism, which is currently work in progress \cite{cyril}.

\section{Conclusion and Discussion}

Ever since the early days of particle cosmology, it has been an ambition to connect particle physics with the early universe directly, without unneccessary degrees of freedom and fine-tunings.  This work is a step in that direction by identifying a toy Higgs-Electroweak sector of the Standard Model as the cause of cosmic inflation, as well as possibly providing a self-contained preheating mechanism. On the other hand, this electroweak inflationary model has the virtue of resolving the Higgs naturalness problem and how a time-dependent (cosmological) electoweak symmtery breaking mechanism can end inflation and reheat the universe.

There are some directions and unanswered questions to be pursued.  First, the periodic potential that is posited would benefit from microphysical explanation.  Such potentials can arise from instanton effects due to a pseudo-Nambu-Goldstone boson, such as a strongly coupled Cooper pair of quarks (i.e. the pion).  Little Higgs models are a natural setting for such constructions and it will be interesting to relate this model to those general classes of electroweak symmetry breaking \cite{Georgi,Arkani-Hamed,Katz}.  Second, we initiated some salient features of the cosmological perturbations in this model, in the context of preheating.  It will be important to do a detailed analysis of cosmological perturbations in this model and to make contact with CMB physics to a greater degree of precision.  Finally, models of this sort produce chiral gravitational waves which leave imprints of TB cross-correlations in the CMB as well as leptogenesis \cite{lwk,aps,Caldwell,Adshead2,Adshead3}. It would be interesting to revisit these mechanisms in the context of these specific models since the scales are now in the TeV range.

\section{Acknowledgements}
  I especially thank Michael Peskin, Matt Strassler, Evan McDonough, Humberto Gilmer and Cyril Creque-Sarbinowski for useful feedback and discussions.  I also thank Subodh Patil, Tucker Manton, Tatsuya Daniel, and Devin Walker for discussions and looking at a draft of this paper.

\bibliographystyle{bibstyle} 
\bibliography{bibo}

\end{document}